# Harmonic-seeded resonant Raman amplification in strong-field ionized nitrogen molecules

Jinping Yao[1], Wei Chu[1], Chuanshan Tian[2], Ziting Li[1,3], Zhanshan Wang[3], and Ya Cheng[1,4,5*]

[1]*State Key Laboratory of High Field Laser Physics, Shanghai Institute of Optics and Fine Mechanics, Chinese Academy of Sciences, Shanghai 201800, China*
[2]*Department of Physics, Fudan University, Shanghai 200433, China*
[3]*School of Physics Science and Engineering, Tongji University, Shanghai 200092, China*
[4]*State Key Laboratory of Precision Spectroscopy, East Normal University, Shanghai 200062, China*
[5]*Collaborative Innovation Center of Extreme Optics, Shanxi University, Taiyuan, Shanxi 030006, China*

[*]*ya.cheng@siom.ac.cn*

**Abstract**

Generation of free-space laser-like emissions of high spatiotemporal coherence and narrow bandwidth in atmosphere opens promising opportunities for remote spectroscopic sensing. Here, we report on generation of such laser-like emissions, which results from the combined contributions of perturbative and non-perturbative nonlinear optical effects in nitrogen molecules exposed to intense mid-infrared laser fields. We systematically investigate the dependence of the generated free-space laser spectrum on wavelength and power of the driver laser. It is revealed that the free-space laser is produced by resonant Raman amplification of the fifth harmonic of the driver pulses in rotational wavepacket of the molecular nitrogen ions.

Nonlinear optical phenomena can be ubiquitously observed in light-matter interactions as far as the strength of electric field of the light is comparable to the interatomic electric fields [1]. Specifically, when focusing intense near- or mid-infrared (mid-IR) laser pulses into gaseous atoms or molecules, nonlinear optical effects can be dramatic which are featured by rapid multi-photon excitation or tunnel ionization of the gaseous media [2] or efficient generation of coherent photons of which the frequency covers an extremely broad spectral range [3,4]. The extreme nonlinearity is mainly a result of the ultrashort temporal duration of the intense driver pulse, which prevents most neutral atoms or molecules from photoionization in the rising edge of the pulse until the arrival of the crest of the pulse where the field strength suddenly reaches its maximum. This unique nature has opened the new area of non-perturbative nonlinear optics, which is also coined as extreme nonlinear optics [5,6].

It is well known that in the perturbative nonlinear optics, both non-resonant and resonant processes can be readily observed. Typically, the latter process is more efficient than the former through resonant enhancement. In contrast, observation of the resonant nonlinear optical processes in the non-perturbative nonlinear regime has rarely been reported. Indeed, in high harmonic generation (HHG) simultaneously driven by a strong IR field and a weak ultraviolet (UV) field, resonant excitation of atoms to an excited state by single-photon absorption of the UV photons has given rise to significant enhancement of HHG efficiency [7-10]. Nevertheless, owing to the weak strength of the UV field, resonant excitation through simultaneous absorption of multiple UV photons has not been reported yet in these experiments. One likely exception is a recent observation reported by M. Chini et al. which may be attributed to HHG in the resonant condition [11], but the physical origin is still under debate [12]. As a matter of fact, in the simple-man model, which has been proven to be successful in the tunneling regime, resonant multi-photon excitation of atoms or molecules to a high-energy level is usually excluded as such process is considered to be insignificant [13]. Until now, search of resonant nonlinear effects in the extreme nonlinear light-matter interactions remains largely unexplored.

Here, we report an observation of resonant Raman amplification in an ionized molecular nitrogen system which is initially prepared by exposure of neutral nitrogen molecules to intense mid-IR fields in the non-perturbative nonlinear regime. The wavelength of the driver laser was tuned away from resonances of the neutral nitrogen molecules so that the interaction is in the non-perturbative regime (i.e., the Keldysh parameter $\gamma=\sqrt{I_P/2U_P}<1$, $I_P$ ionization potential, $U_P$ ponderomotive energy [2]); whereas the fifth harmonic of the driver pulses was near the resonance of ionized nitrogen molecules between the ground and the second excited electronic states. The co-existence of neutral molecules and the molecular ions in the intense fields gives rise to intriguing nonlinear optical effects that has received insufficient attention in the ultrafast optics community so far.

In addition to that incentive, the current investigation will also shed light on the comprehensive physical mechanism of laser-like coherent emissions generated from the molecular nitrogen ions with strong laser fields [14-27]. In particular, our recent investigation on the wavelength dependence of the lasing action in the nitrogen molecular ions indicates that with the 800-nm driver fields, population inversion between the ground and excited states is generated as a result of ultrafast nuclear dynamics after the photoionization of nitrogen molecules. However, with the mid-IR driver fields whose wavelengths range from 1200 nm to 2000 nm, both our experimental observations and theoretical simulations suggest that the population inversion does not exist [28]. Therefore, the mechanism behind the laser-like emission observed with the mid-IR driver pulses is yet to be clarified.

The experiments were carried out using an optical parametric amplifier (OPA, HE-TOPAS, Light Conversion Ltd.), which was pumped by a commercial Ti:sapphire laser system (Legend Elite-Duo, Coherent, Inc.). The OPA enables to generate wavelength-tunable mid-IR laser pulses in the spectral range from 1200 nm to 2400 nm. The mid-IR driver pulses were then focused into the gas chamber filled with air or argon at 1atm by a plano-convex lens to form a plasma channel. A variable attenuation filter was placed before the focal lens to control the laser pulse energy. The output signal was collected by an integration sphere that is connected to the fiber bundle followed by a grating spectrometer (Shamrock 303i, Andor). To examine the evolution of the laser-like emission at ~391 nm as a function of the propagation distance in the plasma channel, we truncated the plasma channel at different locations using a pair of uncoated glass plates. After two reflections from the surfaces of the glass plates, the laser intensity was reduced by nearly one order of magnitude, which leads to the termination of the plasma channel after the second glass plate. Truncation of the plasma channel was achieved by translating the focal lens mounted on a motorized translation stage. Details of the technique can be found in our previous work [29]. The signal was focused into the spectrometer installed with a slit, and the width of the slit was intentionally set to be sufficient for ensuring that the signal could be completely collected during the translation of the focal lens for truncating the plasma channel.

First, we compare the behaviors in growth of the laser-like emissions along the plasma channels produced by the 800-nm and 1910-nm driver pulses, as shown in Fig. 1(a) and (b), respectively. The 800-nm pulses with a pulse energy of 5.6 mJ were focused by an $f$=50 cm lens to produce a single filament in air, resulting in efficient white light generation of which the ultra-broad spectrum covers the transition wavelength from $N_2^+(B^2\Sigma_u^+, v'=0)$ state to $N_2^+(X^2\Sigma_g^+, v=0)$ state (i.e., ~391 nm). When switching to the mid-IR driver pulses, a focal lens with a shorter focal length $f$=25 cm was used because of the much weaker energy of the 1910-nm pulses ( <0.8 mJ ). Thus, tighter focusing was necessary to produce a sufficiently high peak intensity for photoionizing the nitrogen molecules in air.

The spectral feature along propagation in Fig. 1(a) and (b) shows obvious contrast. The laser-like narrow-bandwidth emission in Fig. 1(a) is observed well behind the spectrum of the supercontinnum extends to 391 nm. With our previous investigation, the laser-like emission at ~391 nm produced by the 800-nm driver field can be attributed to seed amplification in the population-inverted molecular nitrogen ions, where the seed is provided by the supercontinnum generated in the plasma channel [28]. At the early stage (i.e., indicated as the stage I in Fig. 1(a)), the intensity of the seed grows linearly with the propagation distance in the plasma channel, as in this stage the emission generated in the forward direction is dominated by the spontaneous emission from the ions in the excited state. Later, when the intensity of the seed reaches the threshold for lasing (i.e., indicated as the stage II in Fig. 1(a) when the stimulated emission begins to dominate), the laser emission at ~391 nm undergoes a rapid growth because of the high gain in the plasma channel [17].

However, the same laser-like emission at ~391 nm in Fig. 1(b) appears immediately when the supercontinnum is spectrally broadened to 391 nm. The sharp difference in Fig. 1(a) and 1(b) indicates that the mechanisms behind the generation of 391 nm emission with the 800 nm and mid-IR (i.e., 1910 nm) driver pulses are different. It also indicates that the coherent emission of the light at 391 nm depends largely on wavelength of the pump pulse. Thus, we measured the emission spectrum with the pump wavelength varying in the range of 1650 nm and 2000 nm, as presented in Fig. 2. For all the pump wavelengths chosen for the experiment, the average power of the laser was maintained at 660 mW. Figure 2(a) shows the spectra of the fifth harmonic generated in air with an $f$=15cm lens as a function of the wavelength of the driver field. As expected, the center wavelength of the fifth harmonic gradually shifts toward red part of the spectrum when the wavelength of the driver field increases. Remarkably, in a wide range of wavelengths of the driver pulses, strong laser-like emissions can be observed at either ~357 nm or ~391 nm superimposed on the spectra of the fifth harmonic. The wavelengths of the coherent emissions at ~357 nm and ~391 nm correspond to the transitions of $N_2^+(B^2\Sigma_u^+, v'=1) \rightarrow N_2^+(X^2\Sigma_g^+, v=0)$ and $N_2^+(B^2\Sigma_u^+, v'=0) \rightarrow N_2^+(X^2\Sigma_g^+, v=0)$, respectively.

To understand how the generated narrow-bandwidth laser-like emissions interplay with the fifth harmonic generation, we performed the same set of wavelength-dependent measurements in 1-atm argon gas, as shown in Fig. 2(b). Here, we chose argon as a reference gas because it has a similar ionization potential and nonlinear coefficient as that of nitrogen molecules but no resonances in the current spectral range. In principle, the harmonic spectra generated from the two kinds of gases should be of little difference. It should be pointed out that the wavelengths of the driver fields were all calibrated using the central wavelengths of fifth harmonic waves generated in argon, which are indicated with black dot-dashed lines in both Fig. 2(a) and (b) for clarity. Comparing Fig. 2(a) with 2(b), one can see the following important features: (1) The two strong peaks at ~357 nm and ~391 nm completely disappear in argon; (2) It is particularly interesting that the laser-like emission is stronger when the peak of the fifth harmonic spectrum is shifted to the blue side of the transition wavelength of 391 nm, whereas in the case that a 800-nm pump laser was used, the laser emission is the strongest when the spectral peak of the seed pulses exactly overlaps the transition wavelength at 391 nm [21].

Since the major difference between the spectra in Fig. 2(a) and (b) exists near the wavelength of driver fields at 1930 nm, we examined the fifth harmonic spectra generated in air and argon as a function of the laser power at this wavelength, as shown in Fig. 3(a) and (b), respectively. When the driver laser power increases, the laser-like emission at ~391 nm generated in air, which is superimposed on top of the broad fifth harmonic spectrum, becomes stronger with the increase of the pump power. Interestingly, we can observe that the generation of laser-like emission at ~391 nm is accompanied by the formation of a "hole" on the blue side of the fifth harmonic spectrum, as shown in Fig. 3(a). The "hole" in the fifth harmonic spectrum indicates strong absorption or optical loss in this wavelength range. In contrast, the fifth harmonic generated in argon shows persistent spectral broadening when the pump power increases from 320 mW to 500 mW, where neither the spectral "hole" nor the narrow-bandwidth peak was observed. When the pump power is above 500 mW, the fifth harmonic spectrum remains unchanged, as shown in Fig. 3(b), which should be a result of intensity clamping in the plasma channel.

Furthermore, we compared the signal intensities at 391.2 nm in air and argon as a function of pump powers, as shown in Fig. 3(c). This wavelength is indicated by the dashed pink lines in Fig. 3(a) and (b), intensity of the former is obviously much stronger. Figure 3(c) shows that the minimum pump power for generating the laser-like emission at 391.2 nm wavelength in air is 450 mW. Afterwards, the generated signal undergoes a rapid growth with the increasing pump power. However, the fifth harmonic signal from argon at 391.2 nm wavelength saturates at 450 mW, indicating that the intensity clamping effect in the light filament occurs near 450 mW at which the nonlinear self-focusing is balanced by the plasma defocusing. It is therefore reasonable to link the generation of the 391-nm laser emission with the initiation of photoionization of nitrogen molecules. To examine the evolution of spectral "hole" with the increasing pump laser power, we plot the differentiated spectrum between air and argon (i.e., $\Delta S = S_{Air} - S_{Ar}$) as a function of driver laser powers in Fig. 3(d). Clearly, when the power of the driver laser reaches 400 mW, the spectral "hole" starts to appear on the blue side of the laser line at ~391 nm. With the increase of the driver laser power, the spectral "hole" becomes broader and its peak continuously shifts toward blue side. Importantly, we notice that the spectral "hole" in Fig. 3(d) overlaps the R-branch rotational spectrum of the nitrogen molecular ions, which provides a hint on the mechanism of the lasing action at ~391 nm as discussed below.

Based on the above systematic experimental results, we come to the following physical picture of combined perturbative and non-perturbative nonlinear optical effects in nitrogen molecules under strong IR laser field. As illustrated in Fig. 4(a), the nonlinear propagation of the strong mid-IR pulses in air initiates nonlinear self-focusing far before the pulses arrive at the geometrical focus. Since air is a centrosymmetric medium, the interaction of the self-focused laser pulses with neutral air molecules will generate odd-order harmonics such as the third and fifth harmonics before the photoionization occurs. When the driver pulses further propagate toward the geometric focus, the peak intensity continuously grows resulting from the linear geometrical focusing and the nonlinear self-focusing. Eventually, the peak intensity of the mid-IR pulses reaches the level that is sufficient to generate air plasma by photoionization, and the spatial collapse of pulses is arrested. At this stage, the molecular nitrogen ions are produced, which are mainly populated in the ground electronic state, as the tunnel ionization dominantly occurs from the highest occupied molecular orbital [30]. Specifically, when the wavelength of the driver pulses is tuned to 1930 nm as chosen in our experiment, the spectrum of the fifth harmonic will cover the transition wavelength of the ionized nitrogen molecular (i.e., ~391 nm). In such case, interaction of the fifth harmonic with the molecular nitrogen ions will efficiently occur through two resonant Raman processes, namely, a resonant Stokes and a resonant anti-Stokes process as illustrated in Fig. 4(b) and (c). The probabilities of the Stokes and anti-Stokes processes are proportional to

$$N_{J-1}\left|\mu_{J-1}^{J'}\right|^2\left|\mu_{J'}^{J+1}\right|^2, \qquad (1)$$

and

$$N_{J+1}\left|\mu_{J+1}^{J'}\right|^2\left|\mu_{J'}^{J-1}\right|^2, \qquad (2)$$

, respectively [31]. Here, $N_{J-1}$ and $N_{J+1}$ denote respectively the populations on the $J-1$ and $J+1$ states of $N_2^+(X^2\Sigma_g^+, v=0)$. $\mu_{J-1}^{J'}$ is the transition dipole moments from $J-1$ state of $N_2^+(X^2\Sigma_g^+, v=0)$ to $J'$ state of $N_2^+(B^2\Sigma_u^+, v'=0)$, whereas $\mu_{J'}^{J+1}$ is the transition dipole moments from $J'$ state of $N_2^+(B^2\Sigma_u^+, v'=0)$ to $J+1$ state of $N_2^+(X^2\Sigma_g^+, v=0)$. Based on the fact that $\left|\mu_{J-1}^{J'}\right|=\left|\mu_{J'}^{J-1}\right|$ and $\left|\mu_{J+1}^{J'}\right|=\left|\mu_{J'}^{J+1}\right|$, the intensity ratio of Stokes and anti-Stokes Raman scattering is determined by the population ratio as follow:

$$I_{Stokes}/I_{anti-Stokes} = N_{J-1}/N_{J+1}. \tag{3}$$

Since the emitted photons from the Stokes process can be absorbed by the counterpart anti-Stokes process and vice versa, the Stokes Raman scattering will prevail if the population in $J-1$ state is higher than that in $J+1$ state, as shown in Fig. 4(b). On the contrary, the anti-Stokes Raman scattering will be stronger than Stokes process if the population in $J+1$ state is higher than that in $J-1$ state, as shown in Fig. 4(c).

It is known that in the thermal equilibrium condition, the rotational wavepacket of neutral nitrogen molecules has a Boltzmann distribution. Generally speaking, after photoionization, the distribution of the rotational wavepacket of the molecular nitrogen ions can still be approximately described with the Boltzmann distribution. Therefore, for the molecular nitrogen ions, the population in a specific rotational $J$ state is expressed by

$$N_J \approx \frac{N_0 hcB_e}{kT}(2J+1)e^{-hcB_e J(J+1)/kT}, \tag{4}$$

, where $N_0$ is total number of the ions in $N_2^+(X^2\Sigma_g^+, v=0)$ state, the rotational constant of the state $B_e = 1.92 \text{cm}^{-1}$, $T$ rotational temperature [32]. At the room temperature (i.e., 300 K), the peak of Boltzmann distribution is $J_{peak}$=7. Meanwhile, in the driver laser fields, which align the nitrogen molecules and molecular nitrogen ions through the cascade rotational Raman process, the peak of Boltzmann distribution will shift toward higher $J$ values as a result of increasing rotational temperature. Meanwhile, the Boltzmann distribution will be broadened as well according to Eq. (4).

Combining Eq. (3) and Eq. (4), the Stokes process will be more efficient than the anti-Stokes process if $J \geq J_{peak}$, which leads to the loss in the R-branch and gain in the P-branch of the fifth harmonic spectrum as observed in Fig. 3(a). With the Boltzmann distribution in Eq. (4), the molecular ions will mainly populate in the states whose rotational quantum numbers $J \geq J_{peak}$ but not $J < J_{peak}$ states. This ensures that the Stokes process will dominate which is consistent with our observation. Furthermore, at the relatively low driver laser powers, the rotational temperature is low and therefore, the value of $J_{peak}$ is small. In this case, the center of the spectral "hole" will be close to the 391 nm lasing line and the width of the "hole" is small. With the increase of the driver laser power, the molecular nitrogen ions will be forced to populate in higher $J$ states with a broader distribution of $J$ values as a result of the increasing rotational temperature, thus the "hole" in the R-branch spectrum becomes broader and its peak shifts to the bluer side. Finally, when the driver laser power is increased to 600 mW, we observe in Fig. 3(a) that the spectral "hole" in the R-branch spectrum almost disappears. We attribute this to the fact that at such a high power of the driver laser, the generated fifth harmonic signal is too intense to be obviously quenched by the resonant absorption from the photoionized nitrogen molecular ions, as the total amount of the ions is ultimately limited by intensity clamping in the plasma channel. Therefore, the results in Fig. 3 agree well with the explanation based on the resonant Raman amplification mechanism.

In addition, we have also observed that the laser-like emission at ~391 nm will disappear when the spectrum of the fifth harmonic of the driver laser shifts beyond the spectral range of the R-branch of the rotational spectrum of the nitrogen molecular ions, as shown in Fig. 2(a). It is now easy to understand because our observation relies on conversion of the R-branch photons into the P-branch photons, which requires the existence of photons in the R-branch of the spectrum.

To conclude, we experimentally investigate the generation of laser-like narrow-bandwidth emission from nitrogen molecular ions in intense mid-IR laser fields. We observe that the laser-like emission is confined in a narrow spectral range, which corresponds to the P-branch of the rotational spectrum of molecular nitrogen ions. The generation of the laser-like emission at ~391 nm is always accompanied by a strong absorption in the R-branch of the spectrum. This characteristic suggests that the laser-like emissions generated with the mid-IR driver lasers can be attributed to resonant Raman amplification. The nonlinear conversion of broad-bandwidth harmonics of ultrashort pulses into narrow-bandwidth coherent sources at various wavelengths opens possibility of remote atmospheric sensing with ultra-high spectral resolutions.

This work is supported by the National Basic Research Program of China (Grant No. 2014CB921303), and National Natural Science Foundation of China (Grant Nos. 11134010, 61575211, 11304330, 61405220 and 11404357).

**Captions of figures:**

Fig. 1 (Color online) (a) The evolution of the supercontinuum generated by 800-nm, 5.6-mJ pump laser in the plasma channel. The data were an average of 5 measurements conducted in the same experimental condition. (b) The evolution of the fifth harmonic generated by ~1910-nm, 0.8-mJ pump laser in the plasma channel. The white arrow indicates the propagation direction of the driver field.

Fig. 2 (Color online) The dependence of the fifth harmonic spectrum generated in (a) air and (b) argon on the driver laser wavelengths. The black dot-dashed line indicates the central wavelength of the fifth harmonic waves generated in argon.

Fig. 3 (Color online) The fifth harmonic spectra generated in (a) air and (b) argon as functions of the power of driver pulses at the wavelength of ~1930 nm. (c) The intensity of the fifth harmonic from air (red circles) and argon (blue squares) at 391.2 nm as a function of the power of the driver laser. (d) The differentiated spectrum of the fifth harmonics generated in air and argon at different powers of the driver laser.

Fig. 4 (Color online) Conceptual illustration of the physical mechanism. (a) In region I, the interaction of the laser pulses with neutral air molecules will generate odd-order harmonics such as the third and fifth harmonics. Due to the geometrical focusing and nonlinear self-focusing, molecular nitrogen ions are produced by photoionization in region II. Rotational Raman amplification is induced through the interaction of the fifth harmonic with the rotational wavepackets of molecular nitrogen ions, resulting in the laser-like coherent emission at ~391 nm. Stokes and anti-Stokes Raman scattering dominates for the case of (b) $N_{J-1} > N_{J+1}$ and (c) $N_{J-1} < N_{J+1}$, respectively.

**Fig. 1**

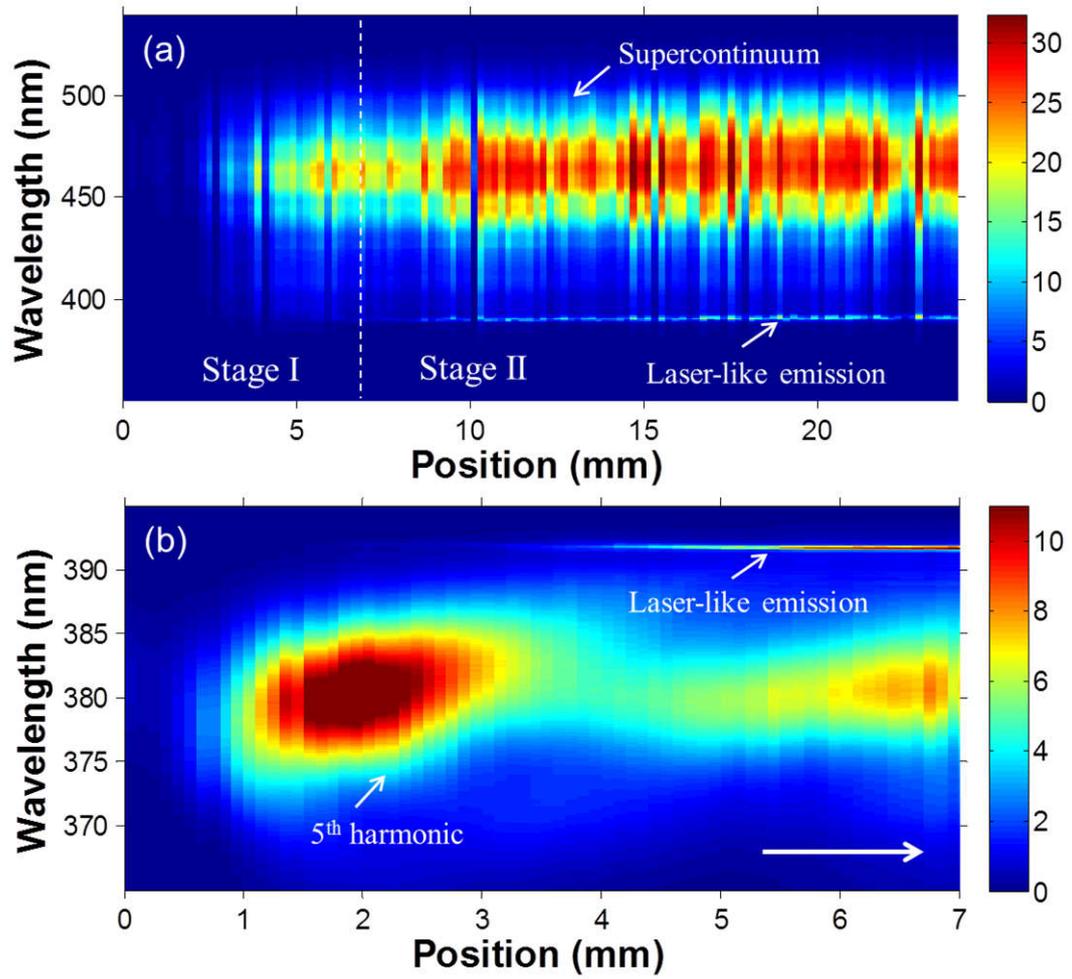

**Fig. 2**

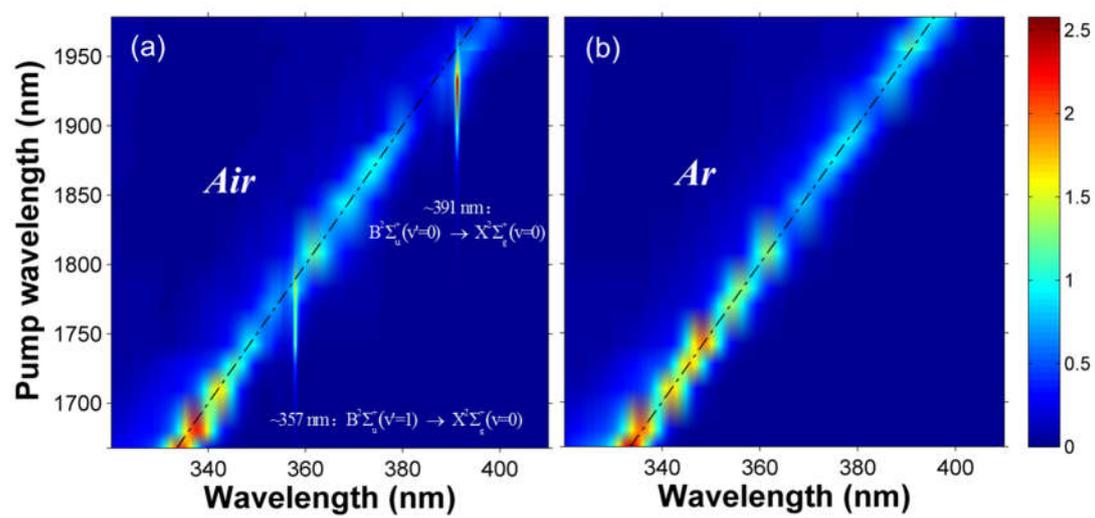
18

**Fig. 3**

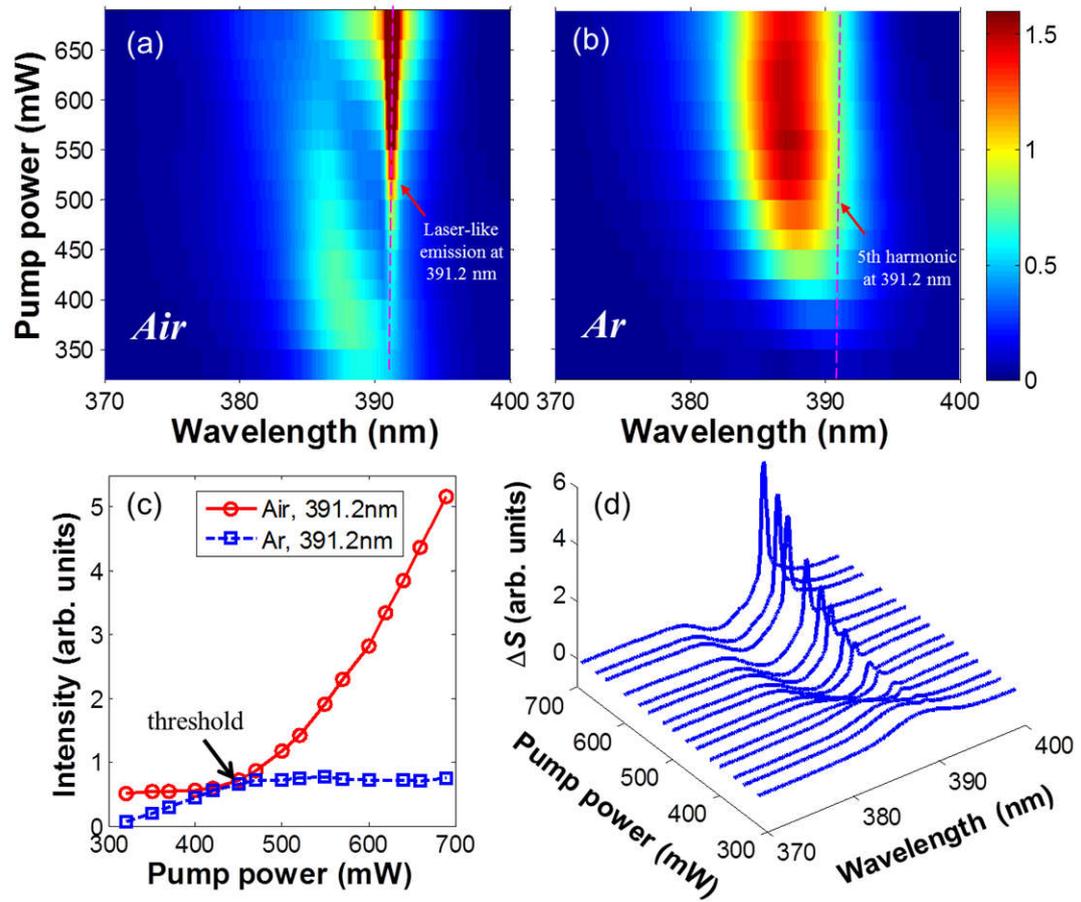

**Fig. 4**

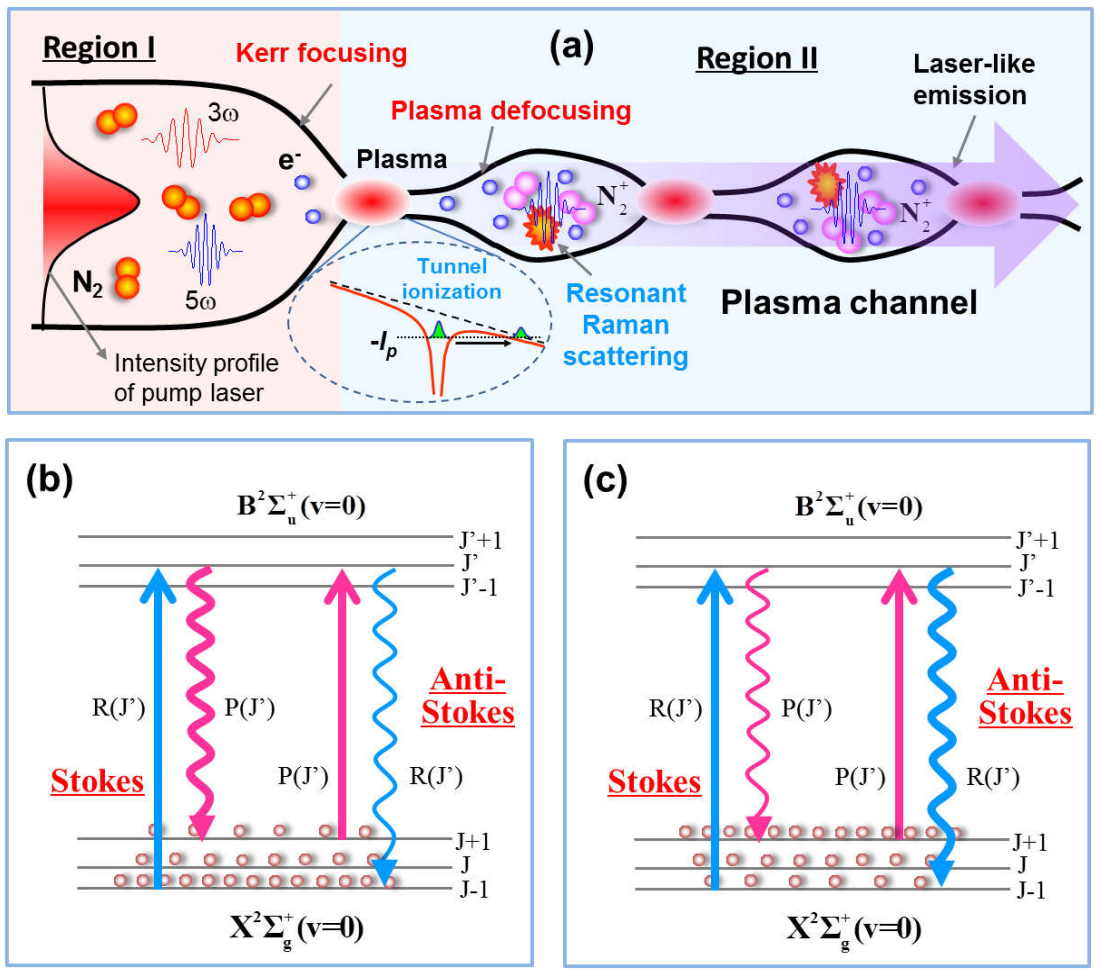